\documentclass[11pt]{article}
\usepackage[utf8]{inputenc}
\usepackage{cite}
\usepackage{listing}
\usepackage{listings}
\usepackage{graphicx}
\usepackage{longtable}
\usepackage{latexsym}
\usepackage{booktabs}
\usepackage{array}
\usepackage{longtable}
\usepackage{lipsum}
\usepackage{chngpage}
\usepackage{setspace}
\usepackage{xcolor}
\usepackage[format=plain,font=small,labelfont=bf]{caption}
\usepackage{float}
\usepackage{framed}
\usepackage{epsfig}
\usepackage{tabularx}
\usepackage{floatflt}
\usepackage{wrapfig}
\usepackage{makeidx}
\usepackage{multirow}
\usepackage[top=2.0cm, bottom=2.0cm, left=2.0cm, right=2.0cm]{geometry}
\usepackage[export]{adjustbox}
\usepackage[english]{babel}
\usepackage[section]{placeins}
\usepackage[toc,page]{appendix}
\usepackage[affil-it]{authblk}
\graphicspath{{figures/}}

\begin{document}

\title{\Large \bf Simulation studies for source optimization in $^{96}$Zr $\beta$ decay}
\date{}

\renewcommand\Authfont{\fontsize{12}{13}\selectfont}
\renewcommand\Affilfont{\fontsize{10}{9.8}\itshape}

\author[1]{S. Thakur}
\author[2]{V. Nanal \thanks{nanal@tifr.res.in}}
\author[1]{P.P. Singh}
\author[1]{R.G. Pillay}
\author[3,4]{H. Krishnamoorthy}
\author[3,4]{A. Mazumdar}
\author[2]{A. Reza}
\author[1]{P.K. Raina}
\author[3,4]{V. Vatsa}

\affil[1]{Department of Physics, Indian Institute of Technology Ropar, Rupnagar Punjab - 140001, India}
\affil[2]{Department of Nuclear and Atomic Physics, Tata Institute of Fundamental Research, Mumbai - 400005, India}
\affil[3]{India-based Neutrino Observatory, Tata Institute of Fundamental Research, Mumbai - 400005, India}
\affil[4]{Homi Bhabha National Institute, Anushaktinagar, Mumbai - 400094, India}

\maketitle

\section*{Abstract}
The single $\beta$ decay of $^{96}$Zr to the ground state of $^{96}$Nb is spin forbidden and poses a great experimental challenge. The $\beta$ decay of $^{96}$Zr can be studied via coincident detection of de-exciting gamma rays in $^{96}$Mo, which is the end product of $^{96}$Nb $\beta$ decay. Simulations are done with four HPGe detector setup ($\sim$\,33\,\% relative efficiency each) to optimize the source configuration. The results  suggest that $\sim$\,70\,g of 50\,\% enriched $^{96}$Zr will yield  sensitivity comparable to the reported results.

\section{Introduction}
\begin{figure}[h!]
\begin{center}
\includegraphics[width=9.0cm,height=4.6cm, trim=5 44 17 1, clip]{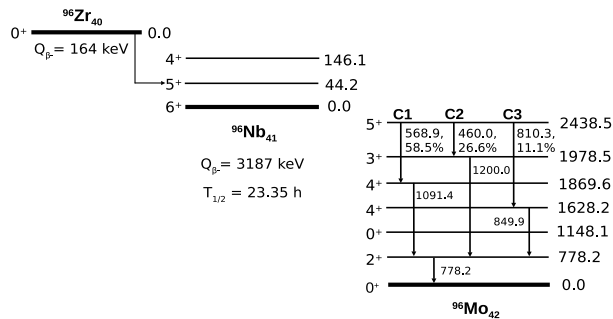}
\end{center}
\caption[Nb decay scheme]{A schematic representation of $\beta$ decay in $^{96}$Zr and $^{96}$Nb (energy values are in keV)~\cite{nndc}.}
\label{Nbdecay}
\end{figure}

The nucleus $^{96}$Zr is one of the two double $\beta$ decay (DBD) candidates, where single $\beta$ decay is spin forbidden and competes with  $\beta\beta$ decay. For $^{96}$Zr, Q$_{\beta\beta}$ is 3.35\,MeV and reported limit for T$_{1/2}$ is 3.1$\times10^{20}$\,yr~\cite{finch_2015} from DBD to excited states of $^{96}$Mo.
A schematic representation of $\beta$ decays of $^{96}$Zr and $^{96}$Nb is shown in Fig.~\ref{Nbdecay} together with prominent gamma decay cascades. There have been several attempts to measure the half-life for $^{96}$Zr $\beta$ decay~\cite{Arpesella_1994, barabash_1996, finch_2016} and the best limit is given as $T_{1/2}\geq 6.2\times10^{19}$\,yr~\cite{Adam_2018}.

Given the relatively small natural isotopic abundance of $^{96}$Zr (2.8\,\%), one of the major challenges in a rare $\beta$ decay study is to improve the sensitivity, which primarily involves the reduction of background to achieve a better signal to noise ratio. Recently, an improved lower limit for T$_{1/2}$ of DBD of $^{94}$Zr to excited states of $^{94}$Mo has been reported using low background setup TiLES~\cite{neha_Zr2017}. In the present work, a feasibility study of the $^{96}$Zr $\beta$ decay through $^{96}$Mo gamma ray cascade using a low background setup of four detectors is carried out employing the coincidence technique for background reduction. 
A setup of four identical HPGe detectors, with the relative efficiency of about 33$\%$, arranged in a plane, is considered in the present study (see Fig.~\ref{detgeom}). The source foil configuration and mounting geometry are optimized for maximizing the coincidence detection efficiency. The $^{96}$Zr is considered to be distributed in $^{nat}$Zr matrix. The results are compared with coincidence measurements of Finch \textit{et al.}~\cite{finch_2016} with a two detector setup. 

\section{Simulation and  Analysis}
\begin{figure}[h!]
\centering
\includegraphics[width=5.6cm,height=3.5cm, trim=100 60 100 50, clip]{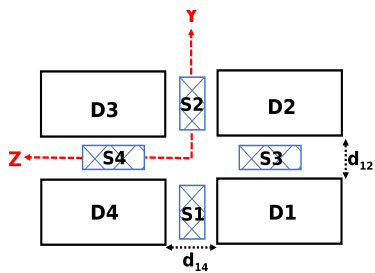}
\caption[Geometry]{A schematic view of source - detector configuration employed in simulation.}
\label{detgeom}
\end{figure}

A simulation program 4HPGeSim has been developed using the GEANT4 (v10.05)~\cite{geant}. Fig.~\ref{detgeom} shows source and detector configuration. The geometry of 4 HPGe detectors is taken to be similar to that of the CRADLE detector at TIFR~\cite{modelling}, having carbon fiber housing and 0.9\,mm thick front window. The source is taken to be $^{nat}$Zr and effect of isotopic enrichment are taken care of by appropriately scaling the number of events generated with the desired fraction ($f$) while retaining the natural material properties for the source. 

Two randomly oriented gamma rays from a chosen cascade (see Fig.~\ref{Nbdecay}) are generated at a given vertex, which is uniformly distributed within the source and detected in the HPGe detectors.  The dimensions of box-shaped source plates (consequently, the  source mass) and their positioning are varied to find the optimum configuration to maximize mass efficiency (M${\epsilon}_c$) -  the product of the source mass and the coincidence photopeak efficiency. The total energy deposited in each detector (E$_{dep}$) is folded by a Gaussian function to account for the detector resolution and the energy detected (E$_{det}$) is recorded. 
Simulation outputs are stored and analyzed in ROOT~\cite{root} for $\sim$\,1\,M events. The photopeak area (N$_i$) is extracted by fitting a sum of Gaussian peak with a quadratic background in singles and coincidence spectra. The efficiency $\epsilon_i$ of detector $i$ is given by
\begin {equation}\label{eq:1}
\epsilon_i =N_i/N_{gen},
\end {equation}
where, $N_{gen}$ is total generated events.  The coincidence counts ($N_{coin}$) in the region of interest are extracted from the two dimensional correlation plots of E$_i$ vs E$_j$ (see Fig.~\ref{568-1091keV_set}(a)). 
The net coincidence counts $N_{c,ij}$ are obtained after proper background and chance correction. The coincidence efficiency for D1-D4 and D1-D2 sets are computed as 
\begin {equation}\label{eq:2}
 \epsilon_{c,ij} = N_{c,ij}/N_{gen},
\end {equation} 
Further,  $\gamma_i-\gamma_j$ ($\gamma_i$ in D4 and $\gamma_j$ in D1) and $\gamma_j-\gamma_i$ ($\gamma_j$ in D4 and $\gamma_i$ in D1) combinations are taken into account while defining total coincidence efficiency $\epsilon_c$.
In a rare decay experiment, the net expected event rate is often quoted in terms of M$\epsilon_{c}$, defined as 
\begin {equation}\label{eq:3}
 M\epsilon_{c} = fM_0\epsilon_c,
\end {equation} 
Where $M_0$ is the total mass of the source, $f$ is the isotopic  fraction.
It should be pointed out that with increasing thickness, the attenuation of emitted gamma rays within the source becomes increasingly important. Hence, the source geometry needs to be optimized to maximize M$\epsilon_{c}$. 
Initially, M$\epsilon_{c}$ is optimized for a two detector setup  D1-D4 (front source) and D1-D2 (side source). For the front source, thickness $t$ is varied, keeping the cross-sectional area of ($l\times w$) constant. 
For the side source, both thickness $t$ and width $w$ are varied, keeping $l$ constant and the effect of variation of $l$ is  investigated separately. Distance between detectors $d_{12}/d_{14}$ is fixed at $t+10$\,mm. 

\section{Results and discussion}
\begin{figure}[h!]
\centering
\includegraphics[width=0.49\linewidth,height=0.44\linewidth]{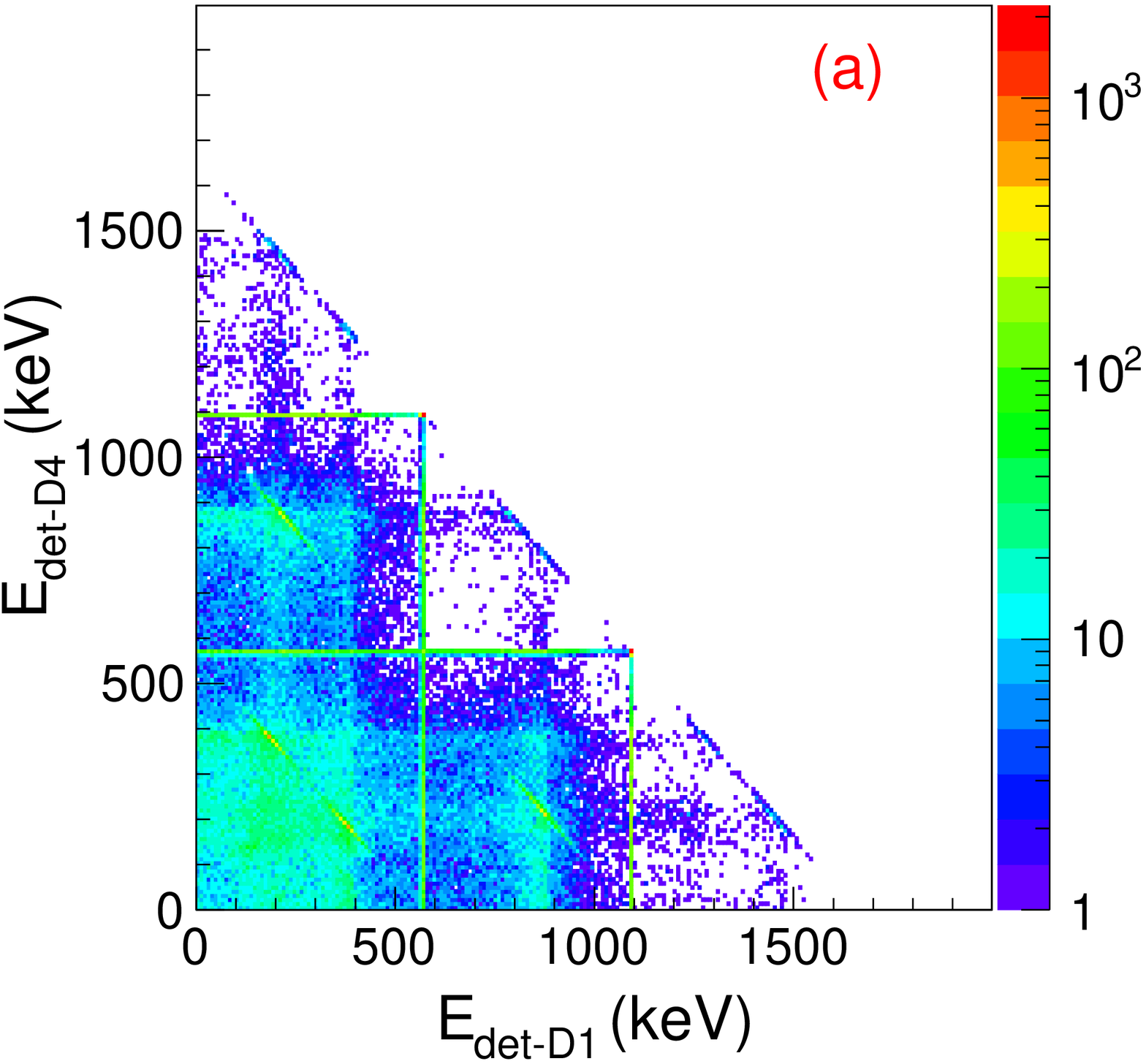}
\includegraphics[width=0.48\linewidth,height=0.45\linewidth, trim=0 196 230 23, clip]{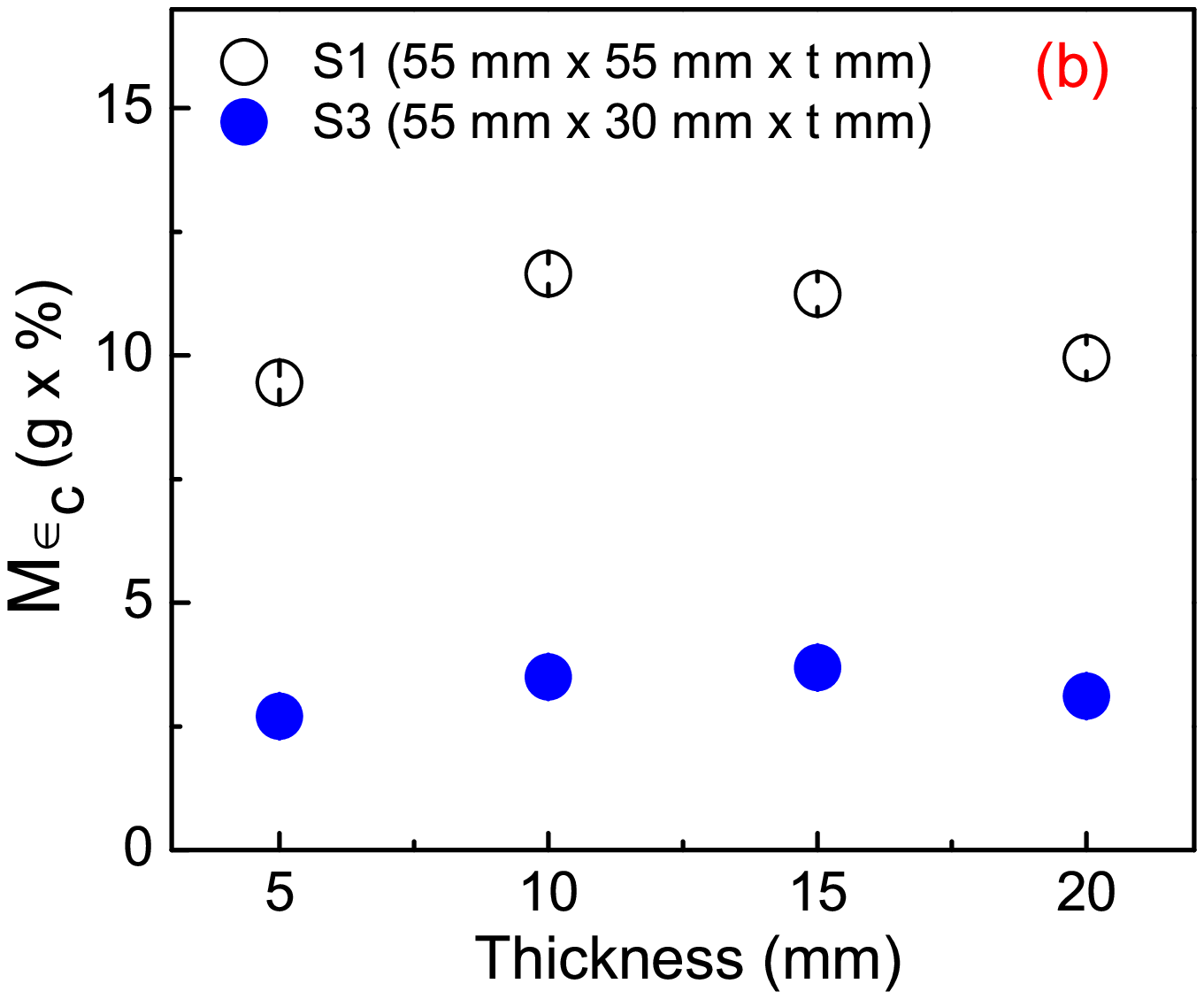}\\
\caption{ (a) Simulated 2D spectrum of the detected energy in D1 and D4, (b) Simulated M${\epsilon}_c$ for 568-1091\,keV gamma rays as a function $t$ for both front (S1) and side (S3) Zr sources.}  
\label{568-1091keV_set}
\end{figure}

Amongst all 3 possible $\gamma-\gamma$ combinations  in the most dominant cascade C1 (see Fig.~\ref{Nbdecay}), 568-1091\,keV  pair is expected to give a cleaner identification of the decay branch. Hence the source geometry optimization has been done for this pair. For side source, it is observed that both singles and coincidence efficiencies show weak dependence on the source width.
As no significant gain in $M{\epsilon}_c$ was observed for $w >30\,$\,mm, $w_{opt}$ is taken to be 30\,mm. The optimal source length ($l_{opt}$) is taken to be 55\,mm same as the crystal length. The M$\epsilon_c$  for 568-1091\,keV gamma pair are plotted as a function of source thickness in Fig.~\ref{568-1091keV_set}(b) for front and side sources.

As expected, for the given mass of the source, the side configuration yields lower efficiency ($\sim$\,half) as compared to the  front source. For 778-1091\,keV pair, as energy is higher, the optimal thickness is somewhat higher than 10\,mm. Although the highest M$\epsilon_c$ is observed for the 568-778\,keV pair, the background in the relevant region will be a crucial factor in the actual experiment.
The optimal source dimensions for 2 detector setup are 55\,mm$\times$55\,mm$\times$10\,mm and 55\,mm$\times$30\,mm$\times$10\,mm for front and side source, respectively. These are used in optimizing the 4 detector setup.
To compare  the present  $M\epsilon_{c}^{opt}$  with the earlier measurement of Finch et al.~\cite{finch_2016}, $M\epsilon_{c}$ is estimated for 568-1091\,keV gamma ray pair for the reference source-detector geometry. Two coaxial HPGe detectors ($\sim$\,88\,mm dia$\times$50\,mm) having 2.54\,mm thick magnesium front window are mounted face to face at a distance ($d$) of 12\,mm. 
A cylindrical source of mass $\sim$\,36.8\,g with $\sim$\,50\,\% enrichment and $\sim$\,60\,mm dia$\times$2\,mm size is considered, giving $M_{ref}\sim$\,18\,g of $^{96}$Zr. As can be seen from Table~\ref{2det}, a significantly large quantity of the Zr source will be needed to achieve M$\epsilon_c$ similar to earlier measurement with present set of detectors. It should be mentioned that measured best reported limit so far on T$_{1/2}$ of $^{96}$Zr employed about 19\,g of ZrO$_2$ powder with 57.3\,\% enrichment for singles gamma ray measurement, resulting in M$\epsilon$ of $\sim$\,37\,g-\%~\cite{Arpesella_1994}.

\begin{table}[h!]
\centering
\caption{A Comparison of M${\epsilon}_c$ (568-1091) for D1-D4 and Ref.~\cite{finch_2016} setup, ($f = 50\,\%$).}
\begin{tabular}{ccccccccc}
\hline
\rule{0pt}{2.5ex} 
Crystal Size  &  Front window       & Source Size   & $d$     & ${\epsilon}_c$ & M    & M${\epsilon}_c$   \\
              & (mm)                &               & (mm)   &   \%     &       (g)            &          (g-\%)           \\
\hline
\rule{0pt}{2.5ex}
88 mm (dia)$\times$50 mm    & 2.54      & 60 mm (dia)$\times$2 mm          & 12     & 0.65    
& 18   & 12       \\
\hline
55 mm (dia)$\times$55 mm   &  0.90   & 55 mm$\times$55 mm$\times$2 mm   & 12     & 0.12    
& 20   & 2        \\
\hline
\end{tabular}
\label{2det}
\end{table} 

For the four detector setup, the total mass is configured in four sources -  S1+S2 (front sources) and  S3+S4 (side sources). Initially, the respective optimum source dimensions obtained in the two detector geometry for the front (55\,mm$\times$55\,mm) and side sources (55\,mm$\times$30\,mm) are employed and sources are positioned symmetrically w.r.t detector crystal for better solid angle coverage.
It may be noted that reducing thickness $t$ from 10\,mm to 5\,mm, permits $d_{14}$ = 7\,mm, which yields $\sim\,$60\,\% gain in $\epsilon_c$. Thus, even though there is a mass decrease of 50\,\%, only $\sim$\,20\,\% decrease is observed in the total M$\epsilon_{c}$. 
The four detector configuration with t = 5\,mm for front and side sources will result in 70\,\% higher M$\epsilon_c$ (see Table~\ref{2det}), but still considerably large  mass $\sim$\,152\,g of $^{96}$Zr will be needed. Hence, further mass optimization needs to be considered. 

As mentioned earlier, dominant contribution comes from sources in the front. So in the first step, only front sources S1 and S2 are employed and the cross-sectional area of the source ($l\times w$) is varied, maintaining  $t_0$ = 5\,mm  and d = 7\,mm to obtain the optimal front source mass (M$_f$). In the second step, a fraction of M$_f$ (30-60\,\%) is distributed as sides sources  S3 and S4. Similar to the first case, $t_0$ and d are kept as 5\,mm and 7\,mm, respectively, and ($l \times w$) is varied keeping M$_f$ fixed ($l<$ crystal length, to avoid edge effects).
The optimal configuration  for t = 5\,mm  is obtained as $M_{eff}\sim$\,72\,g) with M$_{s}\sim 40$\,\% of M$_{f}$. The cross-sectional dimensions ($l_{opt}$, $w_{opt}$)  are 40\,mm$\times$40 mm  for the  front source and 30\,mm$\times$20\,mm for the side source. The M$\epsilon_c (\gamma_1, \gamma_2)$ in Zr matrix with $\sim$\,50\,\% enrichment for the optimal source configuration  are given in Table~\ref{4det_front+side_src1234_all_pair}.  Although higher granularity in the four detector setup is expected to improve the background and reduce the pileup, these effects cannot be quantified at this stage.  

\begin{table}[h!]
\centering
\caption{M$\epsilon_c(\gamma_1, \gamma_2)$ in optimal source configuration ($M_{eff}\sim$\,72\,g) in 4 detector geometry.  }
\begin{tabular}{ccc}
\hline
\rule{0pt}{2.5ex} 
E$\gamma$ (keV)            & $\epsilon_{c}$ (\%)     & M$\epsilon_c$ (g-\%)     \\
\hline
\rule{0pt}{2.5ex} 
\hspace*{-0.22cm}568, 1091   & 0.216    & 15.4    \\
\hspace*{-0.22cm}568, 778    & 0.279    & 20.0    \\
778, 1091                    & 0.172    & 12.3    \\
\hline
\end{tabular}
\label{4det_front+side_src1234_all_pair}
\end{table}  

\section{Conclusion}
Simulation studies are carried out for estimation of M$\epsilon_{c}$ for $\beta$ decay measurement in $^{96}$Zr. The optimization of M$\epsilon_{c}$ is done for four HPGe detector ($\sim$\,33\,\% relative efficiency each) setup with extended sources in a close  geometry for  568-1091\,keV gamma ray pair in the $^{96}$Nb decay cascade. 
It is shown that for $^{96}$Zr $\beta$ decay, even in a four detector configuration, a significantly larger source mass is required to achieve  the reported sensitivity. Present  simulations for a four detector setup show the optimal source configuration  to be  5\,mm thick foils with a cross-sectional area of 40\,mm$\times$40\,mm for front sources and 30\,mm$\times$20\,mm for side sources. This corresponds to  about 72\,g of effective mass with 50\,\% enrichment and can yield M$\epsilon_c$ of $\sim$\,12-20\,g-\% for different gamma ray pairs, which is slightly better than the coincidence measurement reported earlier.   

\section*{Acknowledgement}
This work is supported by the Department of Atomic Energy, Government of India (GoI), under Project No. 12-R\&DTFR-5.02-0300. S. Thakur acknowledges the Ministry of Education, GOI, for PhD research fellowship, and TIFR for supporting the visit related to this work. 


\end{document}